\documentclass{article}
\usepackage[utf8]{inputenc}

\title{An Epidemiological Model for contact tracing with the Dutch CoronaMelder App}

\author{Peter Boncz\footnote{This work was partially financed by ZonMw project 10430022010001.}}
\date{May 2021}

\usepackage{natbib}
\usepackage{graphicx}
\usepackage{amsmath}

\begin{document}

\maketitle

\section{Introduction}
CoronaMelder is the  digital contact tracing app based on exposure notification support in the mobile operating systems of Google and Apple. It was introduced in September 2021 and has been active during the second wave of COVID-19 infections in The Netherlands. This assessment starts with week 42, 2020 (Monday October 12) and ends after week 19, 2021 (Sunday May 16).

In this paper, we quantitatively assess the effectiveness of CoronaMelder and compare its digital contact tracing effectiveness to the manual contact tracing performed by the Dutch health authorities organized in the regional GGD services.
We compute a number of metrics to characterize effectiveness. An important metric is {\bf averted cases}: an estimate of the number of avoided infections of COVID-19, up to May 16. This is not a full total, because this number will continue to rise going forward, because avoided cases in one week will lead to more avoided cases in the next week until the pandemic comes to a full halt. From the number of avoided infections, one can also estimate the amount of {\bf averted deaths} by COVID-19 and {\bf averted hospitalizations} due to COVID-19, both for normal hospital beds and for patients in Intensive Care Units (ICUs) -- under the assumption that each of these three occur for a specific percentage of infections.
We also estimate the {\bf R$_t$ reduction percentage}. We compute $R_t$ mimicking the method of the Dutch health institute RIVM: assuming a 4-day serial interval, i.e. $R_t$ is the growth rate of the amount of daily infected persons with respect to that same amount 4 days earlier.

In this paper, we estimate these metrics for adding manual contact tracing to the standard quarantine and test measures (i.e. "after noting symptoms, quarantine and get a test at GGD immediately"). We also estimate both metrics above for adding the CoronaMelder app to manual contact tracing, under the policy of quarantining immediately after a CoronaMelder warning.\footnote{A Google Sheet with this model can be found on https://bit.ly/2QEwPHc}

\subsection{A Word of Warning}

The computed estimated metrics should not be taken literally. In the first place, both metrics characterize things that did not happen (e.g. "averted infections"). It is fundamentally impossible to prove the validity of such a metric.
In the second place, the estimates are based on models. It is unknown how well these models characterize reality and there are large and unknown variations on their parameters, making the estimated metrics quite uncertain.\footnote{Our estimates could be improved with a parameter sensitivity analysis, though the probability distributions for the parameter values would still be unknown.}

One model is the characteristics of the COVID-19 infection (e.g. infectivity  over time). This model (using a Weibull distribution) is generally used, but it is not exact. In particular, it is known that the disease trajectory varies among patients, something not captured  here. A second model is on the behavior of the Dutch population, in particular, the adherence to measures like "quarantine and test on symptoms", as well as adherence to warnings given to contacts by manual contact tracers and adherence to app warnings. 

However, we have tried to use the best possible information in both models. Regarding the infectivity model, based on publications in scientific literature~\cite{he2020temporal,ashcroft2020covid}. Regarding the adherence, we used information from three interview panels (LISS, PanelClix and the RIVM behavioural studies group). 

Finally, our estimates are not only model-based, as an important part of the numbers that go into our computations are actual amounts of positive tests (which characterize the contact tracing effort accurately) and the actual size of the pandemic in the Netherlands (estimated from hospitalization numbers and serology studies). As such, our estimates are more than just a hypothetical scenario: they estimate the effectiveness of the app and manual contact tracing based on observed quantities of positive tests and hospitalizations in the period October 12 - May 16 in The Netherlands.

Though we believe the literal values of the metrics we estimate should be taken with many grains of salt, we do believe that comparing these metrics between digital and manual contact tracing gives useful insight on the relative effectiveness of these instruments to combat COVID-19.

\subsection{How CoronaMelder Works}

The CoronaMelder app was developed over summer 2020 by a team put together by the Dutch ministry of Health (VWS) and uses the first version of the GAEN API (Google Apple Exposure Notification application programming interface). 
GAEN causes mobile phones to periodically emit random identifiers (numbers) derived from a so-called Temporary Exposure Key (TEK) in bluetooth messages. Each mobile phone generates randomly a new TEK each day. Other devices in the neighbourhood can pick up these messages containing the random identifier.
GAEN does not emit the TEK directly, as this would make tracking easy: the random identifier derived from the TEK changes every 20 minutes.
The app also computes whether multiple of such bluetooth messages received during a span of at least 10 minutes pass a {\em risk threshold}, i.e. have a bluetooth signal strength (pattern) indicating likely proximity within 1.5 meters.

When someone tests positive (the "index" case), the GGD employee performing the manual contact tracing phone call will ask the index whether she uses CoronaMelder, and if yes, will ask whether the index wants to pass a key that can be read from the CoronaMelder app. If this happens, this "GGD key" is sent to the app backend (a central server hosted by VWS) for white-listing.
The user is then asked to (voluntarily) press a button in the CoronaMelder app to also upload the relevant TEKs to that backend, along with the GGD key.
The backend only publishes TEKs from persons with a white-listed GGD key, i.e. only those with a confirmed COVID-19 positive PCR test can upload TEKs.

GAEN remembers received random identifiers from the last 14 days on a mobile phone. To warn people, we typically need less than 14 TEKs: only TEKs for the days the index was infectious.
This "infectiousness window" is determined in the GGD manual tracing phone call, by asking for the day of symptom onset (if the index has no symptoms, the current day is used).
The infectious window starts two days before the onset of symptoms, and ends on the day of the phone call.
So in addition to passing the GGD key to the backend for white-listing, the GGD also passes to the backend the matching infectiousness window. When the CoronaMelder app of the index uploads the 14 TEKs, the backend keeps only the TEKs for the days the index was deemed to be infectious.

All CoronaMelder apps (try to) synchronise six times per day with the backend.
This happens in the background.
When it connects, CoronaMelder downloads all new TEKs and checks whether messages with identifiers generated from these TEKs have been received and, if so, whether these messages exceed the risk threshold.
If this is the case, the CoronaMelder warns the user that there was a dangerous exposure with a COVID-19 infected person on the date those messages were received, and the user is advised to quarantine until 10 days after this dangerous exposure.
From December 1, this advice was changed to also plan a GGD PCR test on the fifth day after the dangerous exposure; if that test returns negative, the user can then stop quarantine immediately.

\section{Modeling Quarantine and Infectivity}

We estimate the effectiveness of manual contact tracing and the CoronaMelder app using a model of the Dutch situation in October 12, 2020 - May 16, 2021 that quantifies how much of the maximum possible infections get realized, under various policies and instruments that influence quarantining behavior.

First we introduce the input to the model (data and parameters) and then introduce the concept of {\bf Tertiary Attack Fraction}, denoted $\lambda$ -- analogous to the weighting function in~\cite{kretzschmar2021isolation}. Then, we discuss the various types of cases Dutch health authority GGD registers every week, and finally we derive specific average $\lambda$ values for each type of case.
Table~\ref{tab:model} contains a compact overview of all parameters and formulas we use for estimating our $\lambda$'s -- these will be discussed and explained in this section.

\subsection{Data Sources and Model Parameters}
We use the following data sources and parameters also shown in Table~\ref{tab:model}:
\begin{itemize}
    \item weekly numbers of PCR tests performed by GGD services for COVID-19 performed, and the amount of positive cases (indexes for contact tracing), as shared in RIVM open data~\cite{rivmopendata}.
    \item weekly reports by RIVM on contact tracing efforts by GGD~\cite{rivmweekly}. The Tuesday afternoon reports mentions the amount of cases found as a result of manual contact tracing. 
    \item weekly reports my the Dutch ministry of health on the effectiveness of the Dutch CoronaMelder app, which mention the amount of people who contacted GGD, mentioning a CoronaMelder warning as one of the reasons for getting a test~\cite{ggd-keys}. Further, the amount of positive cases among these is reported, split out between people who where symptomatic or not when planning the test~\cite{ggd-app-tests}.
    \item NICE, a foundation evaluating ICUs (https://stichting-nice.nl) publishes daily hospitalization statistics, which we extract from Marino van Zelst's github~\cite{zelstcovid}. This number is used to estimate the amount of weekly infections. In particular, serology studies in June 2020 found 5\% of the population with antibodies~\cite{serology}. Noting the 11K hospitalizations (non-ICU) due to COVID-19 up to that point, we estimate the infection hospitalization rate $IHR$=0.0132 and assume this has remained constant. Further, until May 2021 the amount of ICU cases were on average 19\% of hospitalizations, so we estimate the Infection ICU Rate $IIR$=0.0025. Finally, during the second wave (Sept 1, 2020 - May 16, 2021) there were roughly 20K COVID-19 fatalities~\cite{cbs-mortality}, while there were 45K hospitalizations in that period, and hence an estimated 3.4M infections; leading to an Infection Fatality Rate (IFR) of $IFR$=0.006
    \item Panel studies by LISS and PanelClix~\cite{ebbers-cm} and the RIVM behavioral unit\cite{rivmbehavioral}\footnote{We thank Janneke van de Wijgert Wolfgang Ebbers, Ka Yin Leung and Don Klinkenberg for valuable discussions on modeling contact tracing and its  epidemiological parameters.}. In particular, a PanelClix study suggests that $E_{app}$=.58 of people who received a CoronaMelder warning, was never contacted by manual contact tracers. The RIVM behavioral unit makes estimates of quarantine+test adherence to manual contact tracing warnings and CoronaMelder warnings (both $A_{CT}$=$A_{app}$=0.67, being a weighted mix of adherence with and without symptoms)~\cite{klinkenberg-cm}; as well as adherence to quarantine-and-test on symptoms (.50). The latter is down-adjusted to $A_{base}$=.35 under the RIVM assumption that .30 of COVID-19 infections lead to very light to no symptoms (and hence also to no adherence). Finally, RIVM reports $E_{CT}$=.60 of contacts are already alerted by the index personally when GGD CT contacts them.\footnote{RIVM estimates .80 even~\cite{klinkenberg-cm}, but we down-adjust this, as RIVM also states that personal warnings have lower adherance to getting a test than a CT warning (.68 vs .9).}
\end{itemize}

\subsection{Tertiary Attack Fraction $\lambda$}

Let us now explain the calculation of $\lambda$, summarized in Table~\ref{tab:model}. A COVID-19 infected person on average takes $t_{sym}$=5 days to get symptoms, but only after $t_{plan}$=2 more days contacts GGD to plan a test. From that moment it takes $t_{pers}$=1 day to learn about the result, and tell close friends about this personally. However, from planning a test it takes GGD $t_{CT}$=3 days to reach contacts using manual contact tracing. Warnings via the app take a bit less: $t_{app}$=2 days still, mainly because the infected person needs to wait for the contact tracing phone call in order to pass the GGD key.\footnote{Allowing app users who test positively to directly inform app contacts $t_{pers}$=1 day after planning a test, without waiting for the GGD phone call, would improve effectiveness. %However, our modeling here intends to just estimate effectiveness that whas achieved in reality, and does not consider future scenario's.
}

All in all, the typical scenario above, of someone who tests on symptoms, and whose contacts are warned using manual contact tracing, spans $t_{sym}$+$t_{plan}$+$t_{CT}$ =10 days; via the app this spans one day less: $t_{sym}$+$t_{plan}$+$t_{app}$=9 days. Because app adoption $U_{app}$=0.16, only 16\% of indexes has the app and can inform only 16\% of possible contacts\footnote{We ignore any networking effect here for the moment (which could boost app effectivity).} so only $(U_{app})^2$=3\% of contacts is warned by the app. The GGD case numbers show that the manual contact tracing process thoroughly dominates over the app, currently. 

If we look at the infectivity distribution~\cite{he2020temporal} $I_w$, we see that typically no infections happen in the first two days, and most (83\%) happen in days 3-7 of the infection (22\% on day 3, 22\% on day 4, 18\% on day 5, 13\% on day 6 and 8\% on day 7 -- in total 83\% in the first week). As mentioned, the infectivity distribution is known to vary from person-to-person\cite{ashcroft2020covid}; however $I_w$ represents the average trajectory of infectivity over time. We assume that calculating infections using this average trajectory will not come out much different from doing so using probabilistic trajectories (where trajectories of different persons vary around the averages). We keep it simple in this regard, because our further calculation of $\lambda$ would otherwise become much harder to formulate.

Let us assume for a moment a hypothetical basic scenario with just the quarantine-on-symptoms policy in place, i.e. a society without testing, contact tracing or app. Adherence in any scenario is only partial: many people will actually not quarantine. This leads to an adjusted ``adhering'' infectivity distribution $I_a$, where fraction $A_{base}$ no longer infects after reaching the point of taking action because of symptoms (i.e. from day $t_{sym}$+$t_{plan}$=7), but the rest of the population does not quarantine and keeps infecting. Based on average infectivity $I_a(x)$ on day $x$, we also calculate the cumulative fraction of realized infection potential $I_c(x)$, which is the running sum of the former.

Let us go back to a situation where we do have testing, contact tracing and an app.  Given the infectivity distribution, it is clear that an index who decides to test and quarantine after 7 days has already performed 83\% of secondary infections that would have occurred even if testing did not exist. That is, both manual contact tracing and the app can only significantly reduce the number of {\em tertiary} infections. That is why we focus on tertiary infections as the metric of contact tracing effectiveness.

\begin{table}
{\footnotesize\centering
\renewcommand{\tabcolsep}{0.65mm}%
{\renewcommand{\arraystretch}{0.93}%
\begin{tabular}{|c|l|l|l|}
\hline
{\bf symbol} & {\bf value } & {\bf description} & {\bf source}\\
\hline
\hline
$IHR$ & 0.0132 & Infection Hospitalization Rate (excl. ICU hospitalization) %\footnote{Derived from Dutch serology study June 2020: after first wave 5\% seropositive, 11K hospitalizations\cite{serology}}
& NICE~\cite{zelstcovid},\\
$IIR$ & 0.0025 & Infection ICU (Intensive Care Unit hospitalization) Rate
%\footnote{Derived from hospitalization - NICE: cumulative counts ICU are 19\% of regular hospitalization.} 
& Sanquin~\cite{serology} \&\\
{\em IFR} & 0.0060 & Infection Fatality Rate 
%\footnote{Note that RIVM still uses a higher IFR of 1.05 presumably based on first wave statistics. Our IFR is based on hospitalization during the second wave until 31 December 2021, mortality until 1 February 2021, as reported by CBS~\cite{cbs-mortality}.} 
& CBS~\cite{cbs-mortality}\\
\hline
$F_{sym}$ & 0.70 & Fraction of infected people who exhibit symptoms.& RIVM\cite{rivmopendata}\cite{klinkenberg-cm}\cite{rivmbehavioral}\\
$A_{sym}$ & 0.50 & Adherence to q+t (quarantine+test) on symptoms.& RIVM\cite{rivmopendata}\cite{klinkenberg-cm}\cite{rivmbehavioral}\\
$A_{base}$ & 0.35 & effective base adherence $A_{sym}*F_{sym}$& RIVM\cite{rivmopendata}\cite{klinkenberg-cm}\cite{rivmbehavioral}\\
$A_{CT+}$ & 0.75 & q+t adherence to CT warning (with symptoms). & RIVM\cite{rivmopendata}\cite{klinkenberg-cm}\cite{rivmbehavioral}\\
$A_{CT-}$ & 0.50 & q+t adherence to CT warning (no symptoms). & RIVM\cite{rivmopendata}\cite{klinkenberg-cm}\cite{rivmbehavioral}\\
$A_{CT}$ & 0.67 & q+t adherence to CT warning: $F_{sym}A_{CT+}\!+\!(\!1\!-\!F_{sym}\!)A_{CT-}$ & \\
$A_{app+}$ & 0.75 & q+t adherence to app warning (with symptoms). & RIVM\cite{rivmopendata}\cite{klinkenberg-cm}\cite{rivmbehavioral}\\
$A_{app-}$ & 0.50 & q+t adherence to app warning (no symptoms). & RIVM\cite{rivmopendata}\cite{klinkenberg-cm}\cite{rivmbehavioral}\\
$A_{app}$ & 0.67 & q+t adherence to app warning: $F_{sym}A_{app+}\!+\!(\!1\!-\!F_{sym}\!)A_{app-}$ & \\
$E_{CT}$ & 0.60 & Fraction of CT contacts already informed personally by index\ \ & RIVM\cite{rivmopendata}\cite{klinkenberg-cm}\cite{rivmbehavioral}\\
$E_{app}$ & 0.58 & Fraction of app warnings to contacts not reached by  CT\ \  & PanelClix\cite{ebbers-cm}\\  
$F_{CT}$ & 0.50 & Fraction of contacts infected by index found by CT & RIVM\cite{rivmopendata}\cite{klinkenberg-cm}\cite{rivmbehavioral}\\
$F_{app}$ & 0.50 & Fraction of contacts infected by index found by app & RIVM\cite{rivmopendata}\cite{klinkenberg-cm}\cite{rivmbehavioral}\\
$U_{app}$ & 0.16 & Usage of app (estimated from CT phone call data) & GGD\cite{ggd-keys}\\
\hline
$t_{sym}$ & 5 days   & from start of infection to developing symptoms & RIVM\cite{rivmopendata}\cite{klinkenberg-cm}\cite{rivmbehavioral}\\
$t_{plan}$ & 2 days   & from symptoms to planning a test & RIVM\cite{rivmopendata}\cite{klinkenberg-cm}\cite{rivmbehavioral}\\
$t_{CT}$ & 3 days   & from planning a test to contact tracing warnings & RIVM\cite{rivmopendata}\cite{klinkenberg-cm}\cite{rivmbehavioral}\\
$t_{app}$ & 2 days   & from planning a test to app warnings & RIVM\cite{rivmopendata}\cite{klinkenberg-cm}\cite{rivmbehavioral}\\
$t_{pers}$ & 1 day   & from planning a test to personal warnings & RIVM\cite{rivmopendata}\cite{klinkenberg-cm}\cite{rivmbehavioral}\\
\hline
\end{tabular}}\\
\vspace*{1.2mm}
\renewcommand{\arraystretch}{1.2}%
\begin{tabular}{|c|l|l|l|l|l|l|l|l|l|l|l|l|l|l|}
\hline
\multicolumn{15}{|l|}{Weibull infectivity $I_w(x)\!$ =fraction of total infections caused on day $C$ from getting the disease}\\
\multicolumn{15}{|l|}{Adhering infectivity $I_a(x)$=$I_w(x)$ if $x\!\leq\!t_{sym}\!\!+\!t_{plan}$; else $I_a(x)=I_w(x)(\!1\!-\!A_{base}\!)$}\\ 
\multicolumn{15}{|l|}{Cumulative adhering infectivity $I_c(x)$=$\sum_{i=1}^x I_a(i)$}\\
\hline
\hline
day $C$:& 1 & 2 & 3 & 4 & 5 & 6 & 7 & 8 & 9 & 10 & 11 & 12 & 13 & 14..$\infty$\\  
\hline
$I_w(x)$ & 
.0000 & .0000 & .2200 & .2200 & .1800 & .1300 & .0800 & .0500 & .0400 & .0200 & .0200 & .0200 & .0100 & .0000\\
$I_a(x)$ & 
.0000 & .0000 & .2200 & .2200 & .1800 & .1300 & .0800 & .0325 & .0260 & .0130 & .0130 & .0130 & .0065 & .0000\\
$I_c(x)$ & 
.0000 & .0000 & .2200 & .4400 & .6200 & .7500 & .8300 & .8625 & .8885 & .9015 & .9145 & .9275 & .9340 & .9340\\
\hline
\hline
\multicolumn{15}{|l|}{Tertiary Attack Fraction $\lambda(x)$: fraction of infection potential by contacts, index q+t after day $x$}\\
\hline
\multicolumn{15}{|l|}{$\lambda(x)$=$\lambda(1,x)\!:$index does q+t on day $C$ \& warns with 0 latency, all contacts receive+adhere}\\
\multicolumn{15}{|l|}{$\alpha\!\in$\{$CT$,$app$\}:$\lambda_\alpha(x)$: index q+t on day $C$, warns in $t_\alpha$, $F_\alpha$ contacts receive, $A_\alpha$ of them adhere}\\
\hline
$\lambda(1,x)$ &
.0000 & .0000 & .0000 & .0000 & .0484 & .1452 & .2728 & .4092 & .5340 & .6452 & .7125 & .7700 & .8135 & .8491\\
$\lambda_{app}(x)$ &
.0000 & .0000 & .1269 & .2781 & .4259 & .5549 & .6546 & .7254 & .7712 & .8060 & .8338 & .8577 & .8607 & .8607\\
$\lambda_{CT}(x)$ &
.0000 & .0000 & .1511 & .3221 & .4798 & .6085 & .6992 & .7546 & .7977 & .8255 & .8494 & .8577 & .8607 & .8607\\
\hline
\multicolumn{15}{|l|}{$\lambda(l,h)=\sum_{x=1}^{l+h-1} I_w(x)I_c(l\!+\!h\!-\!x)\!:\!$ day $x$ index infects $I_w(x)$ contacts; they infect $I_c(l\!+\!h\!-\!x)$}\\
\multicolumn{15}{|l|}{$\lambda_\alpha(x)= A_{\alpha}F_{\alpha}\lambda(t_\alpha,x)\!+\!(1\!-\!A_{\alpha}F_{\alpha})I_c(x)I_c(\infty)\!:$weighted adherent+reached vs not}\\
\hline
\end{tabular}\\
\vspace*{1.2mm}
\begin{tabular}{|l|l|l|l|}
\hline
\multicolumn{4}{|l|}{All types of cases ($C_\alpha$) counted in week $w$ and their average Tertiary Attack Fractions ($\lambda_\alpha$)}\\
\hline
\multicolumn{4}{|l|}{$C(w)$: all estimated cases=Hospitalization($w$)$/IHR$}\\
\cline{2-4}
$C_{miss}(w)$: missed cases=$C(w)$-$\sum_{\forall \alpha\ne miss} C_\alpha(w)$
& $\lambda_{miss} $ & 0.87 
& $\!$=$I_c(\infty)^2$\\
$C_{sym}(w)$: cases found after symptoms
& $\lambda_{sym} $ & 0.70 
& $\!$=$\!\lambda_{CT}\!(t_{sym}\!\!+\!t_{plan})=\lambda_{CT}(7)$\\
$C_{pers}(w)$: cases found after a personal warning
& $\lambda_{pers} $ & 0.36 
& $\!$=$\!\Lambda_{CT}\!(t_{pers},\!t_{sym}\!\!+\!t_{plan},\!0)$\\
$C_{CT}(w)$: cases found by manual contact tracing
& $\lambda_{CT}   $ & 0.60 
& $\!$=$\!\Lambda_{CT}\!(t_{CT},\!t_{sym}\!\!+\!t_{plan},\!0)$\\
$C_{app+}(w)$: cases found by app with symptoms
& $\lambda_{app+} $ & 0.60 
& $\!$=$\!\Delta(t_{sym}\!-\!1,\!t_{app}\!+\!t_{plan}\!+\!1,\!0)$\\
$C_{app-}(w)$: cases found by app before symptoms
& $\lambda_{app-} $ & 0.27 
& $\!$=$\!\Delta(t_{app},\!t_{sym}\!-\!1,\!t_{app}\!+\!t_{plan}\!\!+\!1)$\\
\hline
\multicolumn{4}{|l|}{$\Delta(l,\!h,\!b)$=$U_{app}\Lambda_{app}(l,\!h,\!b)+(\!1\!-\!U_{app}\!)\Lambda_{CT}(l,\!h,\!b)\!:\!$ weighted $\!\Lambda_\alpha\!\!$ between warnings $\alpha\!\in$\{$CT,\!app$\}:}\\
\multicolumn{4}{|l|}{$\Lambda_\alpha(l,\!h,\!b)$=$\!\sum_{x=1}^{h} \!I_w(x\!\!+\!b)\lambda_\alpha(l\!+\!h\!+\!1-\!x)/\!\sum_{x=1}^{h} \!I_w(x\!\!+\!b)\!:\!$ weighted $\lambda_\alpha$, index loose [$l,l\!+\!h$] days\ \ }\\
\hline
\end{tabular}}
\caption{parameters and equations to model the Tertiary Attack Fraction $\lambda$}
\label{tab:model}
\end{table}
\clearpage

To explain: when a manual contact tracing or app warning reaches an infected contact (i.e., a secondary infection) on day $x$, this person should quarantine and plan a test.
%\footnote{Dutch policy since December 1 2020 says: as soon as possible but not before day 5 or symptom onset -- a negative test then ends quarantine.}
Not all contacts will be reached (fraction $F_{CT}$ resp. $F_{app}$ only) and not all of the ones reached will adhere (fraction $A_{CT}$ resp. $A_{app}$ only) to this policy either.
The secondary contacts who {\em are} reached and adhere will quarantine and stop creating any more tertiary infections after day $x$. Hence, the earlier the warning arrives (the lower $x$ is) the more the tertiary spread is reduced. 
We thus define the {\bf Tertiary Attack Fraction} $\lambda($x$)$ as the fraction of all possible tertiary infections that a primary case causes when contacts get warned on day $x$. This is a number between [0-1], where 1 is the worst-case situation where {\em nobody ever quarantines}.

\subsection{The Various Types of Cases}

GGD gathers weekly counts of various types of cases. In addition, we estimate all new weekly cases (based on hospitalizations), and with that we can also estimate the missed cases. The various case counts for each week $w$ are:
\begin{itemize}
\item $C(w)$: an estimated of the amount of people who were infected with COVID-19 that week. Assuming GGD finds most cases after at least 7 days, and assuming average hospitalization is on day 12, we use the sum of hospitalizations during the 7-day period 5 days later (i.e., hospitalizations in the interval 5 to 12 days from start of week $w$) and dividing that by the IHR=0.0132.
\item $C_{sym}(w)$: cases found by GGD (i.e. people who tested positive) on their own initiative, after developing symptoms.
\item $C_{CT}(w)$: cases found by GGD after manual contact tracing. From the reported numbers of CT (Dutch abbreviation: BCO) cases we subtract fraction $E_{CT}$=.60, because these cases had already been informed personally of their possible infection by the index. That is, these people were already warned, and manual contact tracing was not required for that.
\item $C_{pers}(w)$: the amount of manual contact tracing cases that had already been informed earlier personally by their index, as mentioned under the previous point (i.e. 60\% of the total amount of reported CT cases).
\item $C_{app-}(w)$: pre-symptomatic people who tested positive after an app warning (early cases, without symptoms).\footnote{One would expect $(\!1\!-\!F_{symp}\!)A_{app-}/(\!F_{symp}A_{app+}+(\!1\!-\!F_{symp}\!)A_{app-}\!)$=0.24 of app cases never to develop symptoms. As we see fraction 0.30 of app cases without symptoms when planning a test, only 0.06 are considered pre-symptomatic. Hence we  down-correct to 1/5th of GGD-reported app cases without symptoms as $C_{app-}(w)$ and add the rest to $C_{app+}(w)$.}
\item $C_{app+}(w)$: all other cases found thanks to an app notification.
%We only we only take fraction $E_{app}$=.58 (also for $C_{app-}$ below) to only count people warned by the app, and not by manual contact tracing.

\item $C_{miss}(w)$: all missed cases that week. This is $C(w)$ minus the sum of all categories above (which are exclusive). 
\end{itemize}
\clearpage
Our modeling is based on computing a {\em weekly TAF volume}. This is done by multiplying the average TAF for each case type $\alpha$ by their count:

\begin{equation}
\lambda(w)=\sum_{\alpha \in \{sym,CT,pers,app-,app+,miss\}} \lambda_\alpha C_\alpha(w)
\end{equation}

\subsection{Estimating $\lambda$ per Type of Case}

We now discuss how we estimate the various $\lambda$'s:
\begin{itemize}
\item $\lambda_{miss}$ People who do not test may still quarantine, and hence follow distribution $I_a$. Their realized (fraction of potential) infections through the whole disease is $I_c(\infty$). The secondary infections they cause in turn, are also not on any contact tracing radar and will not be warned; only notice symptoms. Hence they similarly follow the base scenario behaviour and eventually realize $I_c(\infty)$ of their potential infections. Hence we take $\lambda_{miss}$=$(I_c(\infty))^2$=0.87

\item $\lambda_{sym}$ As mentioned, cases found because they tested on symptoms, typically quarantine after 7 days, but it takes up to day 10 for manual contact tracing to reach their contacts. At that moment, 22\% of those secondary infections was infected 8 days ago and thus reached $I_c(8)$=.8652 of their tertiary infection potential. Similarly, another 22\% of contacts was infected 7 days ago and realized $I_c(7)$=.83, 18\% realized $I_c(6)$=.75, 13\% realized $I_c(5)$=.62 and 5\% realized $I_c(4)$=.44. We denote the sum of all this as $\lambda(t_{CT},\!t_{sym}\!+\!t_{plan})$ =$\lambda(3,7)$:
\begin{equation}
\lambda(l,h)=\sum_{x=l}^{l\!+\!h\!-\!1} I_w(x)I_c(l\!+\!h\!-\!x)
\end{equation}
We have to take into account, however, that not all secondary infections warned on day $x$ will be reached and, if so, that they adhere. This means that the non-reached fraction $(1-A_{CT}F_{CT})$ of secondary contagion $I_c(x)$ on day $x$ will be unimpeded and realize $I_c(\infty)$ tertiary contagion:
\begin{equation}
\lambda_\alpha(x)=A_{\alpha}F_{\alpha}\lambda(t_\alpha,t_\alpha\!+\!x)\!+\!(1\!-\!A_{\alpha}F_{\alpha})I_c(x)I_c(\infty)\!
\end{equation}
For simplicity of the formula, we focus only on warnings via manual contact tracing ($\alpha=CT$) and ignore app warnings, because -- as mentioned before -- only 3\% of contacts is covered only by the app due to its low current adoption. All in all $\lambda_{sym}$=$\lambda_{CT}(7)$=0.70.

\item $\lambda_{CT}$ An important goal for manual contact tracing is to find contacts as early as possible, leading to quarantine that is hopefully faster than the 7 days that people who test-on-symptoms achieve. GGD reports for week $w$ on the amount of cases found through manual contact tracing $C_{CT}(w)$, but regrettably does not report a distribution of the time since onset of symptoms for those CT-cases (average moment of infection would be onset-of-symptoms minus $t_{sym}$=5 days).  Hence, we have to estimate this distribution. We use the distribution caused by an index who tested because of symptoms after 7 days, and whose contacts got warned 3 days later (after 10 days). This means the contacts had been ``loose'' for 10-4 days. We thus estimate $\lambda_{CT}$ using a weighted mix of $\lambda_{CT}(x+1)$ with $C$ between low $l$=3 and high $l$+$h$, where $h$=7 (base $b$=0 here):
\begin{equation}
\Lambda_\alpha(l,\!h,\!b)=\!\sum_{x=1}^{h} \!I_w(x\!\!+\!b)\lambda_\alpha(l\!+\!h\!+\!1-\!x)/\!\sum_{x=1}^{h} \!I_w(x\!\!+\!b)
\end{equation}
For $\lambda_{CT}$, we have $\alpha$=$CT$ -- we reuse this formula also for  $\alpha$=$app$ further on. All in all, $\lambda_{CT}$=$\Lambda_{CT}(t_{CT},t\!_{sym}\!+\!t_{plan},0)$ = $\Lambda_{CT}(3,7,0)$=0.60

\item $\lambda_{pers}$ People who get warned personally by an index gain two days (warning time $t_{pers}$=1, whereas $t_{CT}$=3 days). Similar to $CT$ we use $\lambda_{pers}$ = $\Lambda_{CT}(t_{pers},t\!_{sym}\!+\!t_{plan},0)$ =
$\Lambda_{CT}(1,7,0)$=0.36

\item $\lambda_{app+},\lambda_{app-}$ App users tested because of an app warning,  warn their secondary infections both using the app and via manual contact tracing. We therefore mix their $\Lambda$ calculations via $\Delta$ according to app usage $U_{app}$:
\begin{equation}
\Delta(l,\!h,\!b)=U_{app}\Lambda_{app}(l,\!h,\!b)+(\!1\!-\!U_{app}\!)\Lambda_{CT}(l,\!h,\!b)\
\end{equation}
GGD counts people who test because of an app warning separately for those who requested the test while having symptoms ($app+$) and without symptoms yet ($app-$). Without this split, we would estimate $\lambda_{app}$=$\Delta(t_{app},$ $t_{sym}\!+\!t_{plan},0)$=$\Delta(2,7,0)$=0.50, but in order to split it, we assume the cases without symptoms to be in day 3-4 of their infection, and the symptomatic cases to be in days 5-9. 

Here base variable $b$ comes into play and we estimate $\lambda_{app+}$=$\Delta(t_{sym}\!-\!1,\!t_{app}\!+\!t_{plan}\!+\!1,\!0)$=$\Delta(4,5,0)$=0.60 and respectively $\lambda_{app-}$=$\Delta(t_{app},\!t_{sym}\!-\!1,\!t_{app}\!+\!t_{plan}\!+\!1)$ =$\Delta(2,2,5)$=0.27. This low TAF of asymptomatic app cases ($\lambda_{app-}$= 0.27) in comparison with missed cases ($\lambda_{miss}$=0.83) reflects that the promise of GAEN apps to find cases early, can have significant impact on tertiary infections.
\end{itemize}

\section{Estimating Effectiveness}

As defined in equation (1), given the case counts and their respective TAF's, we can compute a {\em weekly TAF volume} $\lambda(w)$.
As TAF volume stands for the amount of infection opportunities, we assume  that it is linearly related with $C(w\!+\!1)$, the number of cases in the next week. Because we can calculate the contribution to weekly TAF volume $C_{CT}(w)\lambda_{CT}$ of manual contact tracing, we can also calculate the additional TAF volume that would have been if manual contact tracing would not exist and these cases would have been missed, i.e. $C_{CT}(w)(\lambda_{miss}-\lambda_{CT}$). This in turn leads to the definition of estimated averted (``saved'') cases $S_{CT}$ in week $w\!+\!1$, $\forall w\geq0$:
\begin{equation}
S_{CT}(w\!+\!1) =  C_{\not CT}(w\!+\!1)*C_{CT}(w)(\lambda_{miss}-\lambda_{CT})/\lambda(w)
\end{equation}
where $S_\alpha$(0)=0, and $C_{\not \alpha}(w)$ is the total amount of cases in week $w$ if $\alpha$ (here, $\alpha$=$CT$) would not have been there. This sequence starts equal to $C$ at $w$=0:
\begin{equation}
C_{\not \alpha}(0)=C(0)
\end{equation}
and continues by cumulatively adding the averted cases to the epidemic, $\forall w\geq0$:
\begin{equation}
C_{\not \alpha}(w\!+\!1) = C_{\not \alpha}(w)*(C(w\!+\!1)/C(w))+S_\alpha(w)
\end{equation}
The averted cases by the app $S_{app}$ is defined similar to $S_{CT}$:
\begin{equation}
S_{app}(w\!+\!1) =  C_{\not app}(w\!+\!1)*
((C_{app-}(w)(\lambda_{mix}\!-\!\lambda_{app-})+C_{app+}(w)(\lambda_{mix}\!-\!\lambda_{app+}))/\lambda(w)
\end{equation}
Here $\lambda_{mix}$=$E_{app}\lambda_{miss}$+$(1\!-\!E_{app})\lambda_{CT}$ because fraction $1\!-\!E_{app}$=.42 of cases found by the app would have been found by  manual contact tracing (only, later), whereas fraction $E_{app}$=.58 of app cases would have been missed altogether. %, so $\lambda_{mix}$ is the weighted average of the TAFs for both cases. 
\vspace*{-2mm}

\begin{table}[h]
{\footnotesize\centering
\renewcommand{\tabcolsep}{0.65mm}%
\renewcommand{\arraystretch}{0.92}%
\begin{tabular}{r|r|r|r|r|r|r|r|r|r|r|r|}
\cline{2-12}
& {\bf w}  & {\bf K$_{app}$} & {\bf C$_{miss}$/C} &{\bf C}  & {\bf C$_{sym}$} & {\bf C$_{pers}$} & {\bf C$_{CT}$} &  {\bf C$_{app-}$} & {\bf C$_{app+}$} & {\bf S$_{CT}$}  & {\bf S$_{app}$} \\
\cline{2-12}
2020-42 & 0 & 8.4 & 0.58 & 134772 & 53531 & 1760 & 1029 & 0 & 344 & 0 & 0\\
2020-43 & 1 & 12.4 & 0.54 & 145302 & 62259 & 2309 & 1243 & 0 & 705 & 344 & 64\\
2020-44 & 2 & 15.6 & 0.53 & 134014 & 55732 & 4284 & 2604 & 0 & 600 & 676 & 173\\
2020-45 & 3 & 15.2 & 0.61 & 114620 & 37278 & 4090 & 2567 & 0 & 380 & 1284 & 239\\
2020-46 & 4 & 13.3 & 0.61 & 97650 & 30140 & 4390 & 2770 & 0 & 372 & 1783 & 260\\
2020-47 & 5 & 13.3 & 0.60 & 91210 & 28481 & 4888 & 3121 & 0 & 328 & 2490 & 305\\
2020-48 & 6 & 13.5 & 0.60 & 83407 & 25070 & 4869 & 3088 & 0 & 377 & 3200 & 332\\
2020-49 & 7 & 12.3 & 0.58 & 100073 & 30195 & 6926 & 4375 & 106 & 447 & 5054 & 479\\
2020-50 & 8 & 13.8 & 0.51 & 122875 & 42627 & 10139 & 6402 & 235 & 567 & 8005 & 728\\
2020-51 & 9 & 14.3 & 0.44 & 138329 & 54945 & 13451 & 8490 & 346 & 718 & 11507 & 1023\\
2020-52 & 10 & 14.4 & 0.55 & 147419 & 44924 & 12976 & 8273 & 251 & 597 & 15524 & 1358\\
2020-53 & 11 & 13.8 & 0.59 & 136131 & 35065 & 12385 & 7942 & 185 & 526 & 17071 & 1431\\
2021-01 & 12 & 12.8 & 0.56 & 113480 & 30728 & 11656 & 7500 & 175 & 433 & 16629 & 1323\\
2021-02 & 13 & 12.3 & 0.63 & 106056 & 23329 & 9345 & 6021 & 147 & 321 & 18161 & 1365\\
2021-03 & 14 & 12.7 & 0.64 & 99465 & 21286 & 8488 & 5485 & 139 & 247 & 19141 & 1379\\
2021-04 & 15 & 12.0 & 0.69 & 94162 & 17572 & 6617 & 4264 & 116 & 210 & 20087 & 1390\\
2021-05 & 16 & 11.4 & 0.71 & 90601 & 15693 & 6374 & 4141 & 76 & 165 & 20882 & 1410\\
2021-06 & 17 & 10.9 & 0.73 & 88707 & 14338 & 5654 & 3670 & 88 & 129 & 21999 & 1433\\
2021-07 & 18 & 10.7 & 0.68 & 88858 & 17127 & 6887 & 4482 & 90 & 151 & 23455 & 1486\\
2021-08 & 19 & 10.0 & 0.65 & 90070 & 18541 & 7608 & 4950 & 95 & 174 & 25579 & 1563\\
2021-09 & 20 & 10.4 & 0.67 & 94463 & 18591 & 7720 & 5018 & 119 & 162 & 28940 & 1705\\
2021-10 & 21 & 10.3 & 0.64 & 105371 & 22556 & 9420 & 6115 & 183 & 171 & 34590 & 1978\\
2021-11 & 22 & 10.5 & 0.63 & 122869 & 26854 & 10831 & 7054 & 136 & 230 & 43354 & 2412\\
2021-12 & 23 & 10.9 & 0.59 & 124990 & 29986 & 12658 & 8238 & 174 & 269 & 47195 & 2542\\
2021-13 & 24 & 10.9 & 0.61 & 123399 & 28490 & 11584 & 7523 & 199 & 235 & 50219 & 2615\\
2021-14 & 25 & 10.5 & 0.58 & 118778 & 28998 & 12097 & 7835 & 241 & 258 & 51608 & 2621\\
2021-15 & 26 & 10.3 & 0.60 & 132791 & 31253 & 12962 & 8427 & 170 & 306 & 61769 & 3074\\
2021-16 & 27 & 10.7 & 0.58 & 133397 & 32710 & 13897 & 9039 & 218 & 275 & 66047 & 3201\\
2021-17 & 28 & 2.9 & 0.55 & 111883 & 29912 & 12301 & 8021 & 149 & 248 & 59076 & 2789\\
2021-18 & 29 & 11.3 & 0.50 & 93551 & 27080 & 11923 & 7772 & 149 & 240 & 52754 & 2413\\
2021-19 & 30 & 10.3 & 0.52 & 74386 & 20363 & 9285 & 6058 & 129 & 160 & 45177 & 1996\\
\cline{2-12}
\multicolumn{4}{r|}{\bf sum:} & 3763515 & 955653 & 269774 & 173518 & 3916 & 10345 & 773599 & 45088\\
\cline{5-12}
\end{tabular}}
\vspace*{-3mm}
\caption{COVID-19 epidemic in The Netherlands from week 42/2020 until week 20/2021. We show weekly cases (C$_\alpha$) of type $\alpha$, estimated averted cases (S$_\alpha$). %K$_{app}$ is the percentage of indexes who indicates to notify contacts using the app.
}  
\label{tab:data}
\vspace*{-4mm}
\end{table}

\subsection{Results}

Table~\ref{tab:data} shows actual measurements of  $C_{sym}$, $C_{pers}$, $C_{CT}$, $C_{app-}$, $C_{app+}$ ; where we clarify once again that the number of manual contact tracing indexes reported by GGD was split by us using fraction $E_{CT}=.6$ in $C_{pers}$ and $C_{CT}$.
The column $C$ is estimated from the hospitalizations reported by NICE, and $C_{miss}$ is  derived from this (by subtracting all GGD-reported cases).
The numbers of averted cases are also estimates, calculated as described earlier in this section. 
Column $K_{app}$ is the percentage of indexes who in the contact tracing phone call indicate to have the app installed and agrees to pass the GGD key that will authorize them to send an app notification. This can be seen as a proxy for app usage ($U_{app}$=.16). The maximum value is 15.6\% but has been going down to slightly above 10\%.\footnote{In week 2021-04, a scandal broke when it was revealed that GGD contact tracing data was being traded illegally via corrupt employees, who had access to too much data with too little oversight.
It seems this cost 16\% of app adoption (K$_{app}$ decreased from 12\% to 10\% by 2021-08), which never fully recovered.}
Studies among GGD personnel~\cite{grip-on-cm} cast doubt on whether this information is properly gathered during all GGD manual contact tracing calls and therefore $K$ should be taken as a lower bound for $U_{app}$. 
The total number of CoronaMelder downloads ($>$4.9M, a fraction .28 of Dutch population) was used as upper bound to arrive at estimate $U_{app}$=.16

We see in the last row that manual contact tracing averted $S_{CT}$=773K infections in this period. This estimate is 4.5\% of Dutch population (17.4M total). For this reason, the effect of saturation is considered to be small ($<$10\%), and we ignore it in the $S_{CT}$ estimate.

Using the sum of estimated averted cases $S_{CT}$ and $S_{app}$, we can also estimate the amount of averted hospitalizations, ICU cases and deaths (by multiplying with resp. $IHR$, $IIR$ and $IFR$).
By calculating the effect of $CT$ and $app$ on the weekly growth rate, and recomputing that into a weekly R$_t$ (growth rate to the power 4/7 -- given the 4-day serial interval Dutch RIVM uses for calculating R$_t$), we can also calculate the average reduction factor on R$_t$ achieved.
The results of such rough calculations are shown in Table~\ref{tab:eval}.

\begin{table}[ht]
{\footnotesize\centering
\renewcommand{\tabcolsep}{0.85mm}%
\renewcommand{\arraystretch}{0.97}%
\begin{center}
\begin{tabular}{|l|r|r|}
\hline
12 October 2020 -& Manual Contact Tracing & CoronaMelder  \\
16 May 2021 & ($\alpha$=$CT$) & ($\alpha$=$app$) \\
\hline
\hline
averted cases S$_\alpha$ & 773599 & 45088 \\
averted hospitalizations & 10212 & 595 \\
averted ICU cases & 1934 & 113\\
averted deaths & 4642 & 271\\
average reduction of R$_t$ & 0.0091 &
0.0005\\
\hline
\end{tabular}
\caption{Estimates of Effectiveness of CT and App as an enhancement of CT. Caveat: unknown error margins, unverifiable estimations. Apply grains of salt!}  
\label{tab:eval}
\end{center}}
\end{table}

\section{Discussion}

Despite the large and unknown error margins of our estimates, it is clear that the effects of the CoronaMelder app have been {\em much} smaller than Dutch manual contact tracing. Our metric of averted cases differs by a factor 17 (i.e., this suggests the app enhanced effectiveness of manual contact tracing by 6\%). 
However, CoronaMelder did have a small but positive effect on the epidemic and it is likely that lives were saved (the rough estimate being a few hundred).

In terms of a qualitative conclusion on the effectiveness, we think one should take into account the following factors that could affect the future picture:

\vspace{2mm}\noindent
{\bf improved app usage} $U_{app}$. We have now estimated app usage at only 16\%.  User studies in LISS and PanelClix indicate, however, that at least 40\% of Dutch population is willing to use the app, but many have not installed it yet. A better communication strategy, and bundling or syndication of the various COVID apps of the Dutch Health Ministry\footnote{Beyond CoronaMelder, there is also CoronaChecker for showing testing/vaccination status, and GGD Contact to self-enter data for manual contact tracing.} thus has fertile ground to significantly increase adoption. Any improvement factor in adoption quadratically improves effectiveness; e.g., at 30\% by a factor 4, and at 40\% by a factor 6.

\vspace{2mm}\noindent
{\bf improved risk score.}
CoronaMelder still uses GAEN version 1 -- while the UK evaluation~\cite{wymant2021epidemiological} which showed stronger effectiveness than we observe here, mentions an increase in effectiveness after adopting GAEN version 2 and using a better risk score method based on Kalman filters~\cite{lovett2020inferring}. Implementing this in CoronaMelder likely will improve $F_{app}$.

\vspace{2mm}\noindent
{\bf more anonymous contacts.}
Unlike the quite successful UK app~\cite{wymant2021epidemiological}, which also operated in environments where at times bars and restaurants were open, Dutch CoronaMelder app has until now only worked {\em during} a lockdown. This means that people reduced the amount of contacts, and that settings (bars, restaurants, universities, etc) with many {\em anonymous} contacts that manual contact tracing cannot easily uncover, did not play much of a role. In a post-vaccination scenario of an opened up society, with a significantly lower R$_t$, manual contact tracing may in fact become less efficacious (due to more anonymous contacts), and the app would in such circumstances become relatively more efficacious (as it can pick up anonymous contacts). The fraction of contacts manual contact tracing can reach could possibly drop by at least a factor 2, while the fraction of contacts reached by the app will not diminish.

\vspace{2mm}
While there is now a disappointing factor 17:1 between infection coverage achieved by manual contact tracing vs. the app, it thus cannot be discounted that in a post-vaccination period during fall/winter 2021-2022 the effectiveness of the app could come much closer to that of manual contact tracing (e.g., 8:6) -- if adoption and the risk scoring were to be improved. This means that CoronaMelder could (partly) compensate for loss in coverage by manual context tracing in an opened-up society.

{\bibliographystyle{abbrv}
\bibliography{references}}
\end{document}